\documentclass[12pt]{article}
\usepackage{amsmath,epsfig,psfrag,cite}

\newcommand{\jtheta}[1]{\vartheta \begin{bmatrix} #1 \end{bmatrix}}
\newcommand{\jstheta}[2]{\vartheta\big[^{#1}_{#2}\big]}
\newcommand{\Z}{{\bf Z}}
\newcommand{\1}{{\bf 1}}

\renewcommand{\Im}{{\rm Im}}

\begin{document}

\title{\vbox{
\baselineskip 14pt
\hfill \hbox{\normalsize TU-843} \\
\hfill \hbox{\normalsize KUNS-2196} } \vskip 2cm
Higher Order Couplings in Magnetized Brane Models \vskip 0.5cm
}
\author{Hiroyuki~Abe$^{1,}$\footnote{email:
 abe@tuhep.phys.tohoku.ac.jp}, \
Kang-Sin~Choi$^{2,}$\footnote{email:
  kschoi@gauge.scphys.kyoto-u.ac.jp}, \
Tatsuo~Kobayashi$^{2,}$\footnote{
email: kobayash@gauge.scphys.kyoto-u.ac.jp} \ \\ and \
Hiroshi~Ohki$^{2,}$\footnote{email: ohki@scphys.kyoto-u.ac.jp
}\\*[20pt]
$^1${\it \normalsize
Department of Physics, Tohoku University,
Sendai 980-8578, Japan} \\
$^2${\it \normalsize
Department of Physics, Kyoto University,
Kyoto 606-8502, Japan} }
\date{}
\maketitle
\thispagestyle{empty}

\begin{abstract}
We compute three-point and higher order couplings 
in magnetized brane models.
We show that higher order couplings are written as 
products of three-point couplings.
This behavior is the same as higher order amplitudes 
by conformal field theory calculations 
e.g. in intersecting D-brane models.
\end{abstract}

\newpage

\section{Introduction}

Extra dimensional field theories, in particular 
string-derived ones, play important roles 
in particle physics and cosmology.
It is one of keypoints how to realize 
four-dimensional chiral theories 
as low-energy effective theories from such 
higher dimensional theories.
Introducing constant magnetic fluxes in extra 
dimensions is one of interesting scenarios 
to realize four-dimensional chiral theories~\cite{Manton:1981es,
Witten:1984dg,Bachas:1995ik,BDL,Blumenhagen:2000wh,
Angelantonj:2000hi,CIM,Troost:1999xn,Ch,Alfaro:2006is}.
Indeed, several models have been studied 
in field theories and string theories.
Furthermore, magnetized D-brane models are 
T-duals of intersecting D-brane models, and 
various interesting models have been constructed 
within the framework of intersecting D-brane models~\cite{BDL,Blumenhagen:2000wh,
Angelantonj:2000hi,Aldazabal:2000dg,
Blumenhagen:2000ea,Cvetic:2001tj}\footnote{
See for a review \cite{Blumenhagen:2005mu} and references therein.}.
Orbifolds with magnetic fluxes and other non-trivial backgrounds 
with magnetic fluxes
have also been studied~\cite{AKO,ACKO,Conlon:2008qi,Marchesano:2008rg}.

In magnetic background, 
zero-modes are quasi-localized  and the number 
of zero modes are determined by a size of 
background magnetic flux.
Such a behavior of zero-modes would be 
important in application for particle phenomenology.
Couplings among those zero-modes 
in four-dimensional effective field theories are obtained 
as overlap integrals of zero-mode profiles in the extra 
dimensional space.
Thus, if they are localized far away from each other 
in the extra dimensional space, 
their four-dimensional couplings would be suppressed and 
such couplings would be useful to explain 
suppressed couplings in particle physics 
such as Yukawa couplings of light quarks and leptons.
Hence, computation of those couplings is quite important.
Indeed, three-point couplings have been calculated 
and their results were found to 
coincide with three-point couplings in intersecting D-brane 
models~\cite{CIM,DLMP}.
(See also \cite{Russo:2007tc}.)
Furthermore, three-point couplings could lead to 
realistic Yukawa matrices.
(See e.g. \cite{ACKO}.)

For further phenomenological applications, 
it is also important to compute higher order couplings.
Indeed, higher order couplings as well as three-point 
couplings have been 
computed within the framework of 
intersecting D-brane models~\cite{CP,AO} and heterotic 
orbifold models~\cite{HV,Atick:1987kd,BKM,Stieberger:1992bj,CK} 
by using conformal field theory (CFT) technique.
Our purpose in this paper is to compute higher order couplings 
in magnetized brane models.
We carry out overlap integrals of three or more wavefunctions  
in the extra dimensional space in order 
to obtain higher order couplings in four-dimensional 
effective field theories.
It will be shown that such higher order couplings 
are written as products of three-point couplings.
This behavior is the same as CFT calculations 
in intersecting D-brane models as well as heterotic 
orbifold models.

This paper is organized as follows.
In section 2, we show our set-up by reviewing 
Ref.~\cite{CIM}.
In section 3, we reconsider the computation of the 
three-point couplings.
Its result have been obtained in~\cite{CIM},  
but here we pay attention to the selection rules 
and rewrite the result, which is convenient to our purpose.
In section 4, we compute the four-point couplings 
and we study its extensions to higher order couplings.
In section 5, we give comments on comparison 
with those couplings in intersecting D-brane models.
Section 6 is devoted to conclusion and discussion.

\section{Set-up}

We consider dimensional reduction of ten-dimensional ${\cal N}=1$
super Yang--Mills theory with $U(N)$ gauge group \cite{SYM}, on a six torus in
Abelian magnetic flux background. 
We factorize the six-torus into two-tori  $(T^2)^3$, each of which is
specified by the complex structure $\tau_d$ and the area 
$A_d = (2 \pi R_d)^2~\Im \tau_d$ where $d=1,2,3$.
{}From the periodicity of torus, the background magnetic flux 
is quantized as \cite{toron}
\begin{equation} \label{toronbg} \begin{split}
 F_{z^d \bar z^d} = {2 \pi i \over \Im \tau_d}  \begin{pmatrix}
   m_1^{(d)} \1_{N_1} & & \\
  & \ddots & \\ & & m_n^{(d)} \1_{N_n} \end{pmatrix}, \quad d=1,2,3,
\end{split} \end{equation}
where $\1_{N_a}$ are the unit matrices of rank $N_a$, $m_i^{(d)}$ are
integers and $z^d$ are the complex coordinates. 
This background  breaks the gauge symmetry 
$U(N) \to \prod_{a=1}^n U(N_a)$ where
$N=\sum_{a=1}^n N_a.$

A magnetic flux in $(4+2n)$ extra dimensions can give rise to chiral
fermions in four dimensions.
Focusing on a submatrix consisting of two blocks,
\begin{equation}
 F_{z^d \bar z^d,ab} = {2\pi i \over \Im \tau_d}
  \begin{pmatrix} m_a^{(d)} \1_{N_a} & 0 \\ 0 & m_b^{(d)} \1_{N_b} 
\end{pmatrix} ,
\end{equation}
the corresponding internal components $\psi_n(z)$ of gaugino fields 
$\lambda(x,z)$ have the form
\begin{equation} \label{KKdecomp}
 \lambda(x,z) = \sum_n \chi_n(x) \otimes \psi_n(z), \quad
 \psi_n(z) = \begin{pmatrix} \psi_n^{aa}(z) & \psi_n^{ab}(z) \\ \psi_n^{ba}(z) &
   \psi_n^{bb}(z) \end{pmatrix},
\end{equation}
where $x$ denotes the coordinates of four-dimensional uncompactified 
space-times, $R^{3,1}$. 
The off-diagonal components of zero-modes of the Dirac equation transform
as bifundamental representations
$\psi^{ab} \sim  \bf (N_a,\overline N_b)$, $\psi^{ba} \sim \bf
(\overline N_a, N_b)$
under $SU(N_a) \times SU(N_b)$, where we omit the subscript 0  
corresponding to the zero-modes, $n=0$.
Since only either of the off-diagonal components has exclusive 
zero-modes, depending on the sign of the relative magnetic flux $M^{(d)}
\equiv m_a^{(d)}-m_b^{(d)}$, the spectrum is chiral; The positive
helicity zero-mode provides $CPT$ conjugate to the one with negative helicity.
With an appropriate gauge fixing, the zero-modes on
each $d$-th $T^2$ are written as \cite{CIM}
\begin{equation} \label{wavefn}
 \psi_d^{j,M^{(d)}}(z^{d}) = N_{M^{(d)}} 
~e^{i \pi M^{(d)} z^{d}{\Im~z^{d}/
   (\Im~\tau_d})} 
~\jtheta{j/M^{(d)} \\ 0}(M^{(d)}z^{d},\tau_dM^{(d)}),
\end{equation}
for $j=1,\dots, |M^{(d)}|$, where the normalization factor $N_M$ is obtained 
as
\begin{equation} \label{normalization}
N_{M^{(d)}} = \left( {2 \Im \tau_{d} |M^{(d)}| \over
   A_{d}^2 } \right)^{1/4} .
\end{equation}
We have the $|M^{(d)}|$ zero-modes labelled by the index $j$.
Note that the wavefunction for $j=k+M^{(d)}$ is identical to one 
for $j=k$.
They satisfy the orthonormal condition,
\begin{equation} \label{orthre2}
 \int d^2z^d\ \psi_d^{i,M^{(d)}}(z^d) \left( \psi_d^{j,M^{(d)}}(z^d)  \right)^* =
  \delta_{ij}.
\end{equation}
The important part of zero-mode wavefunctions is written 
in terms of the Jacobi theta function
\begin{equation} \label{jacobitheta}
 \jtheta{a \\ b}(\nu,\tau) = \sum_{n=-\infty}^\infty \exp\left[ \pi i
   (n+a)^2 \tau + 2 \pi i (n+a)(\nu +b)\right].
\end{equation}
It transforms under the symmetry of torus lattice and
has several important properties \cite{Mu}. 
One of them is the following product rule
\begin{equation} \begin{split} \label{thetaprod}
 \jtheta{{i/M_1} \\ 0}&(z_1,\tau M_1) \cdot \jtheta{j/M_2 \\ 0}(z_2,\tau M_2)
 \\
=& \sum_{m \in \Z_{M_1+M_2}} \jtheta{{i+j+M_1 m \over
     M_1 + M_2} \\ 0 }(z_1 + z_2,\tau(M_1 +M_2)) \\
 & \times \jtheta{{M_2 i - M_1 j + M_1 M_2 m \over M_1 M_2(M_1 +M_2)}
 \\ 0}(z_1 M_2 - z_2 M_1,\tau M_1 M_2(M_1+M_2)). \\
\end{split} \end{equation}
Here $\Z_M$ is the cyclic group of order $|M|$, $\Z_M = \{1,\dots,|M|\}$
where every number is defined modulo $M$.
Although this expression looks asymmetric under the exchange 
between $i$ and $j$, it is symmetric if we take into account the summation.
By using the product property (\ref{thetaprod}), 
we can decompose a product of two zero-mode wavefunctions as follows,
\begin{equation}\begin{split}\label{wvprod}
\psi_d^{i,M_1}(z^d) \psi^{j,M_2}_d(z^d) = &  \frac {N_{M_1} N_{M_2}}{N_{M_1+M_2}} 
\sum_{m \in \Z_{M_1+M_2}}  \psi_d^{i+j+M_1m,M_1+M_2}(z^d)   \\
& \times \jtheta{{M_2 i - M_1 j + M_1 M_2 m \over M_1 M_2(M_1 +M_2)}
 \\ 0}(0,\tau_d M_1 M_2(M_1+M_2)). \\
\end{split} \end{equation}

In this paper, we calculate the generalization of Yukawa couplings to
arbitrary order $L$ couplings
\begin{equation}\label{eq:L-coupling}
 Y_{i_1 \dots i_{L_\chi} i_{L_\chi+1}\cdots i_L} \chi^{i_1}(x) \cdots 
\chi^{i_{L_\chi}}(x) \phi^{i_{L_{\chi}+1}}(x) \dots \phi^{i_L}(x),
\end{equation}
with $L=L_\chi + L_\phi$, where $\chi$ and $\phi$ collectively 
represent four-dimensional components of fermions and bosons, respectively. The system under consideration can be
understood as low-energy effective field theory of open string theory.
The magnetic flux is provided by stacks of D-branes filling in the
internal dimension. The leading order terms in $\alpha'$ are identical to
ten-dimensional super-Yang--Mills theory, whose covariantized gaugino
kinetic term gives the three-point coupling upon dimensional
reduction \cite{CIM,DLMP}. The higher order couplings can be
read off from the effective Lagrangian of the Dirac--Born--Infeld action with
supersymmetrization. The internal component of bosonic and 
fermionic wavefunctions is the same \cite{CIM}.
Therefore it suffices to calculate the wavefunction overlap in the extra
dimensions
\begin{equation} \label{wavefnoverlap}
 Y_{i_1 i_2 \dots i_L} = g_L^{10}  \int_{T^6} d^6z \ 
\prod_{d=1}^3 \psi^{i_1,M_1}_d(z) 
\psi^{i_2,M_2}_d(z) \dots \psi^{i_L,M_L}_d(z),
\end{equation}
where $g_L^{10}$ denotes the coupling in ten dimensions.

\section{Three-point coupling}

In this section, we calculate the three-point coupling 
considering the coupling selection rule. 
As we see later, the three-point coupling provides a building block of higher order couplings.

The gauge group dependent part is contracted by the gauge invariance, 
so that the choice of three blocks $m_a,m_b,m_c$ in (\ref{toronbg}) automatically fixes the relative magnetic fluxes
\begin{equation}
 (m_a - m_b) + (m_b - m_c) = (m_a - m_c), \quad \text{and} 
\quad M_1 + M_2 = M_3,
\end{equation}
where $M_1 = m_a -m_b$, $M_2=m_b-m_c$ and $M_3 = m_a -m_c$.
Here every $M_i$ is assumed to be a positive integer.
This relation is interpreted as the selection rule, in analogy of
intersecting brane case \cite{CIMYukawa,Higaki:2005ie}, to which we come back
later. If it is not satisfied, there is no corresponding gauge
invariant operator in ten dimensions.
In terms of quantum numbers the coupling has the form $\bf (N_a, \overline N_b,1) \cdot (1,N_b, \overline N_c) \cdot (\overline N_a,1,N_c)$ under $U(N_a) \times U(N_b) \times U(N_c)$.

The internal part including the wavefunction integrals 
on the $d$-th $T^2$ gives
\begin{equation}
  y_{ij\bar k} = \int d^2 z \ 
\psi^{i,M_1}(z) \psi^{j,M_2}(z) \left( \psi^{k,M_3}(z) \right)^* .
\end{equation}
The complete three-point coupling is the direct product of those in
$d=1,2,3$ and $g^{10}_3$.
For the moment we neglect the normalization factors $N_M$,
and consider two-dimensional
wavefunctions, omitting the extra dimensional index $d$.
By using the relation (\ref{wvprod}), we can decompose 
the product of the first two wavefunctions 
$\psi^{i,M_1}(z) \psi^{j,M_2}(z)$ in terms of $\psi^{k,M_3}(z) $
and we apply the orthogonality relation (\ref{orthre2}). 
Then, we obtain 
\begin{equation}\label{yijk-1}
 y_{ij\bar k} = \sum_{m \in \Z_{M_3}}  \delta_{i+j+M_1m,k}
 ~\jtheta{{M_2 i - M_1 j + M_1 M_2 m \over M_1 M_2 M_3}
 \\ 0}(0,\tau M_1 M_2 M_3),
\end{equation}
where the numbers in the Kronecker delta is defined modulo $M_3$. 
This expression is symmetric under the exchange 
$(i,M_1) \leftrightarrow (j,M_2)$.

For $\gcd(M_1,M_2)=1$, we solve the constraint from the Kronecker
delta $\delta_{i+j+M_1m,k}$,
\begin{equation} \label{deltaconstraint}
 i+j-k = M_3 l - M_1 m, \quad m \in {\bf Z}_{M_3}, l \in {\bf
   Z}_{M_1}.
\end{equation}
Using Euclidean algorithm, it is easy to see that, in the
relatively prime case $\gcd(M_1,M_2)=1$, there is always a unique solution
for given $i,j,k$. 
This situation is the same as one in intersecting D-brane models 
\cite{CIMYukawa,Higaki:2005ie}.
The argument of the theta function 
in eq.(\ref{yijk-1}) becomes
\begin{equation} \label{thetaarg}
 {M_2 i - M_1 j + M_1 M_2 m \over M_1 M_2 (M_1 + M_2)}
  = {M_2 k - M_3 j + M_2 M_3 l \over (M_3 - M_2) M_2 M_3}.
\end{equation}
Therefore, the three-point coupling is written as 
\begin{equation}  \label{3pt}
  y_{ij\bar k}(l) = \jtheta{{M_2 k - M_3 j + M_2 M_3 l \over M_2 M_3(M_3-M_2)} \\ 0}
  (0,\tau (M_3-M_2) M_2 M_3),
\end{equation}
where $l$ is an integer related to $i,j,k$ through
(\ref{deltaconstraint}).
This is called the 2-3 picture, or the $j$-$k$ picture, 
where the dependence on $i$ and $M_1$
is only implicit.

In the case with a generic value of $\gcd(M_1,M_2)=g$, we can show
\begin{equation} \label{modyukawa}
 y_{ij \bar k} =  \sum_{n=1}^g \vartheta \left[ \begin{matrix}
    {M_2k - M_3j + M_2 M_3 l \over M_1 M_2 M_3 } + {n \over g} \\ 0 \end{matrix}
  \right](0,\tau M_1 M_2 M_3).
\end{equation}
The point is that, for a given particular solution $(i,j,k)$, the
number of general solutions satisfying Eq.~(\ref{deltaconstraint})
is equal to $g$.  We can use a similar argument as above, now considering
${\bf Z}_{M_1/g}$ and ${\bf Z}_{M_3/g}$ instead of the original region.
There is a unique
pair $(l,m)$ in $({\bf Z}_{M_1/g},{\bf Z}_{M_3/g})$ satisfying the
constraint (\ref{deltaconstraint}), i.e. ,
\begin{equation}
 {i+j-k \over g} = {M_3 \over g} l - {M_1 \over g} m.
\end{equation}
Obviously, when $(l,m)$ is a particular solution, 
the following pairs,
\begin{equation} \label{argshift}
 \left(l+ \frac{M_1}{g},m+\frac{M_3}{g} \right) 
\in({\bf  Z}_{M_1},{\bf Z}_{M_3}) ,
\end{equation}
also satisfy the equation with the same right-hand side (RHS).
Since ${\bf Z}_{M_1}$ and ${\bf Z}_{M_3}$ are respectively unions of $g$ identical
copies of ${\bf Z}_{M_1/g},{\bf Z}_{M_3/g}$, there are $g$ different
solutions. 
This situation is the same as one in intersecting D-brane models 
\cite{CIMYukawa,Higaki:2005ie}.
If we reflect the shift (\ref{argshift}) in
(\ref{thetaarg}), we obtain the desired result (\ref{modyukawa}).

There can be Wilson lines $\zeta \equiv \zeta_r + \tau \zeta_i$,
whose effect is just a translation of each wavefunction~\cite{CIM}
\begin{equation}
 \psi^{j,M}(z) \to \psi^{j,M}(z+\zeta), \quad \text{ for all } j.
\end{equation}
Thus the corresponding product for (\ref{thetaprod}) is 
obtained as 
\begin{equation} \begin{split} 
 \jtheta{{i/M_1} \\ 0}&((z+\zeta_1)M_1,\tau M_1) \cdot 
\jtheta{j/M_2 \\ 0}((z +\zeta_2)M_2,\tau M_2)
 \\
=& \sum_{m \in \Z_{M_1+M_2}} \jtheta{{i+j+M_1 m \over
     M_1 + M_2} \\ 0 }((M_1+M_2)(z+\zeta_3),\tau(M_1 +M_2)) \\
 & \times \jtheta{{M_2 i - M_1 j + M_1 M_2 m \over M_1 M_2(M_1 +M_2)}
 \\ 0}(M_1M_2(\zeta_1 -\zeta_2)),\tau M_1 M_2(M_1+M_2)), \\
\end{split} \end{equation}
where $M_3 = M_1 + M_2$ and $\zeta_3 M_3 = \zeta_1 M_1 + \zeta_2 M_2$.

\begin{figure}[h] \begin{center}
\psfrag{i M1}[c]{$i,M_1$}
\psfrag{j M2}[c]{$j,M_2$}
\psfrag{k M3}{$k,M_3$}
\includegraphics[height=3cm]{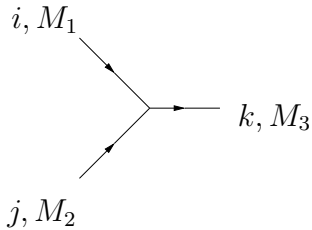}
\caption{A three-point coupling provides a building block of higher
  order couplings. 
This diagram corresponds to the three-point coupling
(\ref{Yukawa}). The direction of an arrow depends on the holomorphicity
of the corresponding external state.}  
\label{f:3pt}
\end{center} \end{figure}

Finally, we take into account the six internal dimensions $T^2 \times
T^2 \times T^2$. Referring to (\ref{wavefnoverlap}), essentially the
full coupling is the direct product of the coupling on each
two-torus. The overall factor in (\ref{wavefnoverlap}) is the physical 
ten dimensional gauge coupling $g_3^{10} = g_{\rm YM}$, since this is
obtained by dimensional reduction of super Yang--Mills theory.
Collecting the normalization factors (\ref{normalization}) from
(\ref{wvprod}), the full three-point coupling becomes
\begin{equation} \label{Yukawa} \begin{split}
 Y_{ij \bar k} = & g_{\rm YM} \prod_{d=1}^3 \left({2 \Im \tau_d \over
   A^2_d}{M_1^{(d)} M_2^{(d)} \over M_3^{(d)}}\right)^{1/4} \\
&\times \exp\left(i \pi (M_1^{(d)} \zeta_1^{(d)} \Im \zeta_1^{(d)} + M_2^{(d)} \zeta_2^{(d)} \Im \zeta_2^{(d)} + M_3^{(d)}
   \zeta_3^{(d)} \Im \zeta_3^{(d)})/\Im \tau_d \right)
\\ &\times \sum_{n_d=1}^{g_d} \vartheta \left[ \begin{matrix}
    {M_2^{(d)}k - M_3^{(d)}j + M_2^{(d)} M_3^{(d)} l \over M_1^{(d)}
      M_2^{(d)} M_3^{(d)} } + {n_d \over g_d} \\ 0 \end{matrix} 
  \right](M_2^{(d)} M_3^{(d)}(\zeta_2^{(d)} - \zeta_3^{(d)}),\tau_d M_1^{(d)} M_2^{(d)} M_3^{(d)}) .
\end{split} \end{equation}
Here the index $d$ indicates that the corresponding quantity is the
component in $d$-th direction. 
For later use, it is useful to visualize the three-point coupling like
Feynman diagram in Fig. \ref{f:3pt}.

\section{Higher order coupling}

\subsection{Four-point coupling}

We calculate the four-point coupling 
\begin{equation} \label{4pt}
 y_{ijk\bar l} \equiv \int d^2 z \ 
\psi^{i,M_1}(z) \psi^{j,M_2}(z) \psi^{k,M_3}(z) 
\left( \psi^{l,M_4}(z) \right)^* ,
\end{equation}
and represent it in various ways.
The main result is that the four-point coupling can be expanded by 
three-point couplings. Thus by iteration, we can generalize it to higher
order couplings.

We consider the case without Wilson lines, since the generalization is
straightforward. The product of the first two wavefunctions 
$\psi^{i,M_1}(z) \psi^{j,M_2}(z) $ in (\ref{4pt}) is the
same as in (\ref{wvprod}). Again, we suppose $M_1 + M_2 + M_3 = M_4$.
Then the product of the first three wavefunctions 
$\psi^{i,M_1}(z) \psi^{j,M_2}(z) \psi^{k,M_3}(z) $
in (\ref{4pt}) gives
\begin{equation} \label{threeprod}
\begin{split}
 \sum_{m \in \Z_{M_1 + M_2}} & \sum_{n \in \Z_{M_4}}
\psi^{i+j+k+M_1 m + (M_1 + M_2) n , M_4}(z)   
  ~\jtheta{{M_2 i - M_1 j + M_1 M_2 m \over M_1 M_2(M_1+M_2)} \\ 0}(0,\tau
 M_1 M_2 (M_1 + M_2)) \\
 & \times \jtheta{{M_3 (i+j+M_1 m) - (M_1 + M_2)k + (M_1+M_2)M_3 n
   \over (M_1+M_2) M_3 M_4} \\ 0}(0,\tau(M_1+M_2)M_3 M_4).
\end{split} \end{equation}
Now, we product the last wave function $\left( \psi^{l,M_4}(z) \right)^*$
 in (\ref{4pt}), acting on the first factor in (\ref{threeprod}), yielding
the Kronecker delta 
$ \delta_{i+j+k+M_1 m + (M_1 + M_2) n,l} $.
The relation is given modulo $M_4$, reflecting that $i,j,k,l$ are
defined modulo $M_1,M_2,M_3,M_4$, respectively. It is non-vanishing if
there is $r$ such that
\begin{equation} \label{selrule}
 i+j+k+M_1 m + (M_1 + M_2) n = l + M_4r.
\end{equation}
We solve the constraint equation in terms of $n$.

For $\gcd(M_1,M_2,M_3)=1$,
any coupling specified by $(i,j,k,l)$ satisfies the constraint.
For a coupling $y_{ijk\bar l}$, fixing $(m,r)$ there is always a
unique $n$ satisfying the constraint.
This means that by solving the constraint equation in terms of $n$,
we can remove the summation over $n$ in (\ref{threeprod}). The result is
\begin{equation} \label{sdecomp-1} \begin{split}
 y_{ijk\overline l} = \sum_{m \in \Z_{M_1+M_2}} & \jtheta{{M_2 i - M_1
     j + M_1 M_2 m \over M_1 M_2 M} \\ 0} (0, \tau M_1 M_2M)
 \cdot \jtheta{{ M_3 l - M_4 k + M_3 M_4 r \over M M_3
     M_4} \\ 0 }(0,\tau M_3 M_4 M ) ,
\end{split} \end{equation}
where $M
= M_1 + M_2 = -M_3 + M_4$. 
This form (\ref{sdecomp-1}) 
is expressed in terms of only `external lines', $i, j, k, l$, 
and in the
`internal line' $r$ is uniquely fixed by $m$ from the relation
(\ref{selrule}).
This is to be interpreted as expansion in terms of three-point
couplings (\ref{3pt}).
{}From the property of the theta function, we have relations like
$y_{ij\bar k} = y_{\bar \imath \bar \jmath k}^*$, etc.
Thus we can write 
\begin{equation} \label{4pt-4}
 y_{ijk\bar l} = \sum_{m \in \Z_{M_1+M_2}} y_{i j \bar m}(m)
 \cdot  y_{k   m \bar l}(r) ,
\end{equation}
where $m$ and $r$ are uniquely related by the relation
(\ref{selrule}). 
Recall that three-point coupling can be
expressed in terms of `two external lines' depending on the 2-3 `picture.'

The result (\ref{sdecomp-1}) can be written by arranging the summation of 
quantum numbers as follows,
\begin{equation} \label{sdecomp-2} \begin{split}
y_{ijk\overline l}  = \sum_{s \in \Z_{M_1+M_2}} & \jtheta{{M_2 s-Mj+M_2 Mr \over
      M_1M_2M} \\ 0}(0,\tau M_1 M_2M)
 \cdot \jtheta{{-Ml+M_4 s + MM_4 n \over M_3 M_4 M} \\ 0}
       (0,\tau M_3 M_4 M).
\end{split} \end{equation}
Here, we rewrite (\ref{selrule}) 
\begin{align}
 i + j + M_1 m &= s + (M_1+M_2) r, \nonumber \\
 -k + l +M_3 r &= s + (M_1+M_2) n,
\end{align}
by introducing 
an auxiliary label $s$, defined modulo
$M=M_1+M_2=-M_3+M_4$. This is uniquely fixed by other numbers from
(\ref{selrule}) and it can be traded with $m$. Thus we arrive at the
second form (\ref{sdecomp-2}), which becomes 
\begin{equation} \label{4pt-4-2}
 y_{ijk\bar l} 
  = \sum_{s \in \Z_{M_1+M_2}}  y_{i j \bar s}
 \cdot  y_{k  s \bar l} .
\end{equation}
The second expression (\ref{sdecomp-2}), explicitly depends on the  
`internal line' $s$. 
It is useful to track the intermediate quantum number $s$. 

\begin{figure}[t] \begin{center}
\includegraphics[height=3cm]{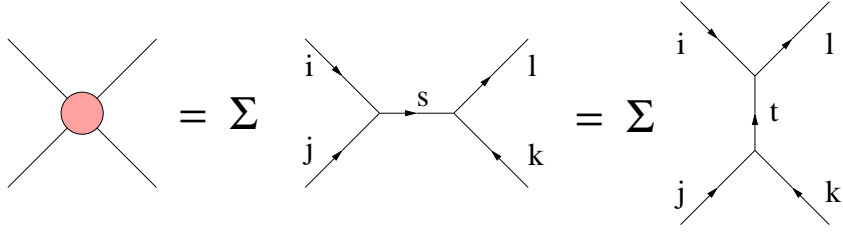}
\caption{A four-point coupling is decomposed into products of
  three-point couplings. It also has `worldsheet' duality. We have
  another `$u$-channel' diagram.}
\label{f:4ptdecomp}
\end{center} \end{figure}

We saw that in the case $\gcd(M_1,M_2)=1$, there is a unique solution.
Since we expand higher order coupling in terms of three-point
couplings, if any of them have degeneracies as in (\ref{modyukawa}),
i.e., $\gcd(M_i,M_j)=g_{ij}>1$, we should take into account their effects.
It is interpreted that each three-point coupling contains a flavor
symmetry $\Z_{g_{ij}}$~\cite{ACKO-2}.
For the four-point coupling with 
$\gcd(M_1,M_2) = g_{12}$ and $\gcd(M_3,M_4)=g_{34}$ we have
also $\gcd(g_{12},g_{34})=g = \gcd(M_1,M_2,M_3,M_4)$, without loss of
generality (see below).
Employing the `intermediate state picture', or the ($j$-$s)\times(s$-$l)$
picture, in the last expression in
(\ref{sdecomp-2}), we have
\begin{equation} \begin{split}
 \sum_{p \in \Z_g} \sum_{s \in \Z_{M_1+M_2}} & \jtheta{{M_2s-Mj+M_2 Mr \over
      (M-M_2)M_2M} + {p \over g} \\ 0}(0,\tau (M-M_2)M_2M) \\
  & \times \jtheta{{-Ml+M_4s + MM_4 n \over M M_4 (M_4-M)} + {p \over g} \\ 0}
       (0,\tau M M_4 (M_4-M)).
\end{split} \end{equation}
It shows that the two symmetries $\Z_{g_{12}}$ and $\Z_{g_{34}}$ are
broken down to the largest common symmetry $\Z_g$, due to the
constraint. Otherwise we cannot put together the vertices with the
common intermediate state $s$.

Reminding that we are examining the overlap of four wavefunctions, and
it {\em does not depend on the order of product}.
If we change the order of the product in (\ref{4pt}), namely consider
the product of the second and the third wavefunctions
$\psi^{j,M_2}(z) \psi^{k,M_3}(z) $
first, we have
differently-looking constraint relation
which is equivalent to (\ref{selrule}) undergoing the decomposition,
\begin{align}
 j + k + M_2 m' &= t + (M_2+M_3) r', \nonumber \\
 -i + l +M_1 r' &= t + (M_2+M_3) n'.
\end{align}
This looks like the `$t$-channel' and we have
\begin{equation} \label{tdecomp} \begin{split}
 y_{ijk\overline l} = \sum_{t \in \Z_{M'}} & \jtheta{{M_3t-M' k+M_3 M' r' \over
      (M'-M_3)M_3M'} \\ 0}(0,\tau (M'-M_3)M_3M')) \\
  & \times \jtheta{{-M'l+M_1t + M'M_1 n \over M' M_1 (M_1-M')} \\ 0}
       (0,\tau M' M_1 (M_1-M')) \\
  = \sum_{t \in \Z_{M'}} & y_{i\bar l t} \cdot y_{jk\bar t},
\end{split} \end{equation}
with $M' = -M_1 + M_4 = M_2 +M_3$.
The result  has a behavior like `worldsheet' {\em duality} in those of
Veneziano and Virasoro--Shapiro \cite{VSV}. This means that, in decomposing the
diagram, the position of an insertion does not matter. 

If we have Wilson lines, we just replace the three-point couplings
by those with Wilson lines (\ref{Yukawa}).

\subsection{Generic $L$-point coupling}
We have seen that the four point coupling is expanded in terms of
three-point couplings.
We can generalize the result to obtain arbitrary higher order couplings.
The constraint relations and the higher order couplings are 
always decomposed into products of three-point couplings. 
It is easily calculated by Feynman-like diagram.

The decompositions (\ref{sdecomp-1}),(\ref{sdecomp-2}),(\ref{tdecomp}) 
are understood as
inserting the identity expanded by the complete set of orthonormal
eigenfunctions
$\{ \psi^{i,M}_n \}$  as follows.
For example, we split the integral  (\ref{4pt}) as 
\begin{equation} \label{split-1}
  y_{ijk\bar l} = \int d^2 z d^2z'\ \psi^{i,M_1}(z)
 \psi^{j,M_2}(z) \delta^2(z-z')
 \psi^{k,M_3} (z') \left(
 \psi^{l,M_4} (z') \right)^*.
\end{equation}
Then, we use the complete set of orthonormal eigenfunctions 
$\{ \psi^{i,M}_n \}$ of the Hamiltonian with a magnetic 
flux $M$.
That is, they satisfy 
\begin{equation}\label{complet-set}
\sum_{s,n} \left( \psi^{s,M}_n  (z) \right)^* \psi^{s,M}_n  (z') 
= \delta^2(z-z').
\end{equation}
We insert LHS instead of the delta function $\delta^2(z-z')$ in 
(\ref{split-1}).
Since $\psi^{i,M_1}(z) \psi^{j,M_2}(z)$ is decomposed 
in terms of $ \psi^{s,M_1+M_2}_n  (z)$, it is convenient to 
take $M = M_1 +M_2$ for inserted wavefunctions 
$\left( \psi^{s,M}_n  (z) \right)^* \psi^{s,M}_n  (z')$.
In such a case, only zero-modes of $\psi^{s,M}_n (z)$ appear 
in this decomposition.
If we take $M \neq M_1 + M_2$, higher modes of $\psi^{s,M}_n (z)$ 
would appear.
At any rate, when we take $M = M_1 +M_2$, 
we can lead to the result (\ref{sdecomp-2}) and 
(\ref{4pt-4}).
On the other hand, we can split 
\begin{equation} \label{split-2}
  y_{ijk\bar l} = \int d^2 z d^2z'\ \psi^{j,M_2}(z)
 \psi^{k,M_3}(z) \delta^2(z-z')
 \psi^{i,M_1} (z') \left(
 \psi^{l,M_4} (z') \right)^*,
\end{equation}
and insert (\ref{complet-set}) with $M=M_2 +M_3$.
Then, we can lead to (\ref{tdecomp}).
Furthermore, we can calculate the four-point coupling after 
splitting 
\begin{equation} \label{split-3}
  y_{ijk\bar l} = \int d^2 z d^2z'\ \psi^{i,M_1}(z)
 \psi^{k,M_3}(z) \delta^2(z-z')
 \psi^{j,M_2} (z') \left(
 \psi^{l,M_4} (z') \right)^*.
\end{equation}
How to split corresponds to `s-channel', `t-channel' and 
`u-channel'.
Note that only zero-modes appear in 
`intermediate states', when we take proper values of $M$ 
because of the product property.

We have considered the four-point couplings 
with $M_1+M_2+M_3 = M_4$ for $M_i >0$.
We may consider the case with 
$M_1+M_2 = M_3 + M_4$ for $M_i >0$, 
which corresponds to 
\begin{equation} \label{4pt-2}
 y_{ij\bar k \bar l} \equiv \int d^2 z \ 
\psi^{i,M_1}(z) \psi^{j,M_2}(z) \left( \psi^{k,M_3}(z) \right)^* 
\left( \psi^{l,M_4}(z) \right)^* .
\end{equation}
In order to consider both of this case and the previous case at the same time, 
we would have more symmetric expression for the four-point coupling
\begin{equation} \label{4pt-3}
 y_{ijkl} = \int d^2 z \ \psi^{i_1,M_1}(\tilde z)
 \psi^{i_2,M_2}(\tilde z)
 \psi^{i_3,M_3} (\tilde z)
 \psi^{i_4,M_4} (\tilde  z) ,
\end{equation}
by defining
\begin{equation} \label{ext}
 \psi^{i,-M}(\bar z) \equiv  \left( \psi^{i,M} (z) \right)^*,
\end{equation}
with
$$M_1 + M_2 + M_3 + M_4 =0, $$
where some of $M_i$ are negative, 
and $\tilde z = z$ for $M >0$ and $\tilde z = \bar z$ for $M <0$.  

\begin{figure}[t] \begin{center}
\includegraphics[height=2cm]{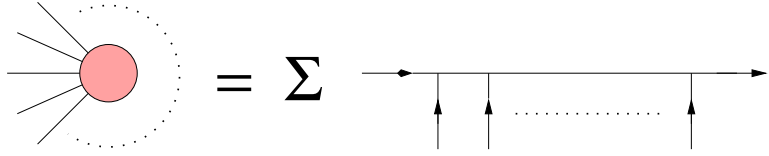}
\caption{Likewise, any amplitude with arbitrary external lines is
  decomposed into product of three-point amplitudes.}
\end{center} \end{figure}

We can extend the above calculation to the $L$-point coupling,
\begin{equation}
 y_{i_1 i_2 \dots i_L} \equiv \int d^2 z \prod_{j=1}^L \psi^{ i_j,M_j} 
(\tilde z), 
\end{equation}
with the extension as in (\ref{ext}). We have then the selection rule
\begin{equation}
 \sum_{j=1}^L M_j = 0,
\end{equation}
where some of $M_j$ are negative.
The constraint is given as
\begin{equation}
 \sum_{j=1}^L \left( i_j + \left(\sum_{l=1}^j M_l \right )r_j \right)
 = 0.
\end{equation}
Again, it shows the conservation of the total flavor number $i_j$,
reflecting the fact that each $i_j$ is defined modulo $M_j$.
We can decompose $L$-point coupling into $(L-1)$ and three-point couplings
\begin{align}
 \sum_{j=1}^{L-3} \left( i_j + \left(\sum_{l=1}^{j} M_l \right )r_j \right) + i_{L-2}
 &= s -  K r_{L-1}, \nonumber  \\
  i_{L-1} + i_L + M_{L-1} r_{L-1} &= - s - K r_{L-2},
\end{align}
where
\begin{equation}
K = \sum_{k=1}^{L-2} M_i = - M_{L-1} - M_L ,
\end{equation}
is the intermediate quantum number.
Therefore if $\gcd(M_1,M_2,\dots,M_L)=1$, by induction we see that there is a unique solution by Euclidean algorithm.
By iteration
\begin{equation} \label{iteration}
 y_{i_1 i_2 \dots i_L} = \sum_s y_{i_1 i_2 \dots i_{L-2} s} \cdot
 y_{\bar s
   i_{L-1} i_{L}} ,
\end{equation}
we can obtain the coupling including the normalization.
Thus, we can obtain $L$-point coupling out of $(L-1)$-point coupling.
Due to the independence of ordering, we can insert (or cut and glue) any node.

\begin{figure}[t] 
\psfrag{a}[r]{$i_1,M_1$}
\psfrag{b}[c]{$i_2,M_2$}
\psfrag{c}[c]{$i_3,M_3$}
\psfrag{d}[c]{$i_4,M_4$}
\psfrag{e}{$i_5,M_5$}
\psfrag{f}[c]{$s_1$,$M_1$+$M_2$}
\psfrag{g}[c]{$s_2$,$M_1$+$M_2$+$M_3$}
\begin{center}
\includegraphics[height=3cm]{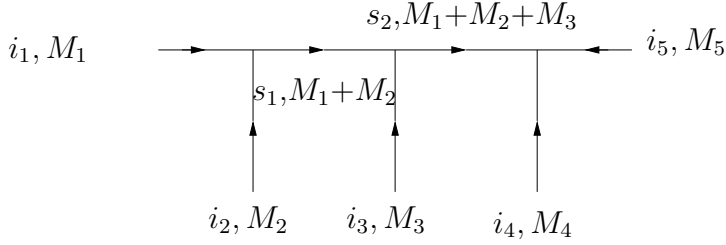}
\caption{Five-point coupling. No more independent Feynman-like diagram
for different insertion.}
\label{f:5pt}
\end{center} \end{figure}

As an illustrating example we show the result for the five-point coupling.
We employ $s$-channel-like insertions, by naming
intermediate quantum numbers $s_i$ as in Fig. \ref{f:5pt}.
We have
\begin{equation} \begin{split}
 y_{i_1 i_2 i_3 i_4 i_5} =& \prod_{j=1}^5
 \jstheta{i_j/M_j}{0}(zM_i,\tau M_i) \\
 =& \sum_{s_1,s_2 }
 \jtheta{{M_2 s_1 - (M_1 +M_2)i_2 + M_2(M_1 +M_2)l_1 \over M_2
     (M_1+M_2)(M_1+2M_2) } \\ 0} (0,M_1 M_2 (M_1 +M_2)\tau)\\
 & \times
  \jtheta{{(M_1+ M_2) i_3 - M_3 s_1 + M_3(M_1 +M_2)l_2 \over M_3 (M_1+M_2)(M_1+M_2+M_3) } \\ 0} (0,(M_1+M_2)M_3(M_1+M_2+M_3))\\
 & \times
  \jtheta{{(M_1+M_2+M_3) i_4 - M_4 s_2  + M_4(M_1 +M_2+M_3)l_3 \over
      M_4 (M_1+M_2+M_3)(M_1+M_2+M_3+M_4) } \\ 0}(0,-(M_4+M_5)M_4 M_5 \tau),
\end{split} \end{equation}
where
$$ s_1 \in \Z_{M_1+M_2}, \quad  s_2 \in \Z_{M_1 + M_2       + M_3}.
$$
{}From the regular patterns of increasing orders, we can
straightforwardly generalize the couplings to arbitrary order.

Now, taking into account full six internal dimensions, as in
three-coupling case (\ref{Yukawa}), we have various normalization factors
besides the product of theta functions. Again, from the product relation of
theta function (\ref{thetaprod}) we have
\begin{equation} \begin{split} \label{symfactor}
 s_L & g_{\rm YM}^{L-2} {\alpha'}^{(L-4+L_\chi/2)/2} \\
\times &\prod_{d=1}^3
\Bigg( {2 \Im \tau_d 
   \over A^2_d} \sum_{M_i^{(d)} > 0}|M_i^{(d)}| \Bigg)^{-\frac14}
\Bigg( {2 \Im \tau_d 
   \over A^2_d} \sum_{M_i^{(d)} < 0}|M_i^{(d)}| \Bigg)^{-\frac14}
\prod_{i=1}^L \left( {2 \Im \tau_d |M_i^{(d)}|
   \over A^2_d} \right)^{\frac14} .
\end{split} \end{equation}
Recall that $L_\chi$ is the number of fermions in the couplings 
(\ref{eq:L-coupling}). 
We have $g^{10}_L = s_L g_{\rm YM}^{L-2}
{\alpha'}^{(L-4+L_\chi/2)/2}$ in (\ref{wavefnoverlap}), where 
symmetric factor $s_L$ comes from higher order expansions of lower-level 
completion of Yang--Mills theory, having also an expansion parameter
$\alpha'$. 
In open string theory, it is 
the Dirac--Born--Infeld action, and it is unknown beyond the quartic
order in $\alpha' F$ \cite{Koerber:2002zb}.
The dependence of ten-dimensional gauge coupling $g_{\rm YM}$ and Regge
slope $\alpha'$ can be easily accounted by order counting \cite{Po}.
Note that $g_{\rm YM}$ is dimensionful.
This factor (\ref{symfactor}) is non-holomorphic in the complex structure $\tau$ and
complexified K\"ahler modulus $\alpha' J = B + i A/4 \pi^2$, where
$B_{z^d \bar z^d}$ is the antisymmetric tensor field component in $d$-th
  two-torus. 
They are interpreted as originating from the K\"ahler potential \cite{CIM,DLMP}.
The product $\prod M_i^{1/4}$ is the leading order approximation of
Euler beta function and its multivariable generalization, which is the
property of dual amplitude.

As an example of full expressions, we show the four-point coupling 
among scalar fields, $Y_{ij\bar l \bar m}\phi^i \phi^j (\phi^l)^*
(\phi^m)^* $, where $\phi^i$ and $(\phi^l)^*$ 
($\phi^j$ and $(\phi^m)^*$) correspond to the magnetic flux 
$M_1^{(d)}$ ($M_2^{(d)}$).
For simplicity, we consider the case with vanishing Wilson lines 
and $\gcd (M_1,M_2)=1$.
The full coupling $Y_{ij\bar l \bar m}$ is obtained as 
\begin{equation}
Y_{ij\bar l \bar m} =  g_{\rm YM}^2\prod_{d=1}^3 \left({2 \Im \tau_d \over
   A^2_d}{M_1^{(d)} M_2^{(d)} \over M_3^{(d)}}\right)^{1/2} 
\sum_{k \in \Z_{M^{(d)}_1+M^{(d)}_2}}  y^{(d)}_{ij\bar k} \ (y^{(d)})^*_{k
  \bar l \bar m} ,
\end{equation}
up to $s_L$, where 
\begin{equation}
y^{(d)}_{ij\bar k} = 
\jtheta{{M^{(d)}_2 k-M^{(d)}j+M^{(d)}_2 M^{(d)}r \over
      M^{(d)}_1M^{(d)}_2M^{(d)}} \\ 0}(0,\tau_d M^{(d)}_1 M^{(d)}_2M^{(d)}).
\end{equation}
This scalar coupling with $s_L=1$ appears from ten-dimensional 
super Yang-Mills theory and satisfies the relation 
$Y_{ij\bar l \bar m} = Y_{ij \bar k} (Y)^*_{k \bar l \bar m}$ 
for the three-point coupling $Y_{ij \bar k}$ in 
eq.~(\ref{Yukawa}).

\section{Intersecting D-brane models}

Here we give comments on the relation between the results 
in the previous sections and higher order couplings 
in intersecting D-brane models, 
i.e. CFT-calculations.



There is well-known $T$-duality relation between magnetized and 
intersecting brane models.
In intersecting brane case, the wavefunctions are highly localized
around intersection points, whereas magnetized brane wavefunctions are
fuzzily delocalized over the entire space.

Under the `horizontal' duality with respect to real axis, $X_z
\leftrightarrow 2 \pi \alpha' A_z$.
The parameter is changed as
\begin{equation}\label{T-dual}
 \tau \leftrightarrow J, \quad \zeta \leftrightarrow \nu .
\end{equation}
Still the translational offset $\nu$ is the Wilson line.
Thus, the magnetic flux gives the slope 
$A_{\bar z}^i = -\frac{i}{2} F_{z \bar z}^i z = \frac{\pi}{\Im \tau}M_i$
and the corresponding quantum number is the
`relative angle,' for small angles,
\begin{equation}
 \pi \theta_{i} = \frac{M_i}{\Im J}.
\end{equation}
The selection rule due to the gauge invariance becomes
\begin{equation}
 M_1 + M_2 = M_3 \leftrightarrow \theta_1 + \theta_2 = \theta_3.
\end{equation}

In the intersecting brane case, as well as heterotic string case,
there have been CFT calculation of higher order
amplitude \cite{Atick:1987kd,AO,CK} using vertex operator insertion \cite{HV,CIMYukawa,CP,BKM}.
There are vertex operators $V_i$ corresponding to massless modes.
We compute their $L$-point amplitude,
\begin{equation}\label{L-amp}
 \langle V_1 V_2 \dots V_L \rangle .
\end{equation}
We have operator product expansion (OPE),
\begin{equation}\label{ope}
V_i(z) V_j(0) \sim \sum_k \frac{c_{ijk}}{z^{h_{ijk}}}V_k(0),
\end{equation}
with $h_{ijk}= h(V_k) - h(V_i) - h(V_j)$,
where $h(V_l)$ is the conformal dimension of $V_l$.
This OPE corresponds to (\ref{wvprod}).
Furthermore, the coefficients $c_{ijk}$ correspond 
to the three-point couplings in four-dimensional effective 
field theory.
In Ref.~\cite{CIM}, it is shown that the above three-point 
coupling $c_{ijk}$ in intersecting D-brane models corresponds 
to the T-dual of the three-point couplings 
$Y_{ijk}$ in magnetized D-brane models.

Now, let us consider the 
$L$-point amplitude $\langle \prod_i V_i(z_i) \rangle $.
We use the OPE (\ref{ope}) to write the $L$-point 
amplitude in terms of $(L-1)$ point amplitudes.
Such a procedure is similar to one 
in the previous sections, where 
we write $L$-point couplings in terms of three-point 
couplings.

For example, the CFT calculations for the four-point 
couplings $c_{ijkl}$ in the intersecting D-brane models 
would lead 
\begin{equation}
c_{ijkl} \sim \sum_s c_{ij \bar s}c_{s kl},
\end{equation}
and
\begin{equation}
c_{ijkl} \sim \sum_t c_{ik\bar t}c_{t jl},
\end{equation}
depending on the order of OPE's, i.e. 
s-channel or t-channel.
Thus, the form of the four-point couplings 
as well as $L$-point couplings $(L>4)$ is almost 
the same as the results in the previous sections.
Note that in eq.(\ref{wvprod}), a product of two wavefunctions 
is decomposed in terms of only the 
lowest modes.
On the other hand, in RHS of Eq.~(\ref{ope}), higher modes 
as well as lowest modes may appear.
However, dominant contribution due to the
lowest modes are the same, because 
$c_{ijk}$ for the lowest modes $(i,j,k)$ 
corresponds exactly to $Y_{ijk}$ for the lowest modes.

Let us examine the correspondence of couplings 
between magnetized models and intersecting D-brane models 
by using concrete formulae.
In the intersecting D-brane models, 
the amplitude (\ref{L-amp}) is decomposed into the classical and the
quantum parts,
\begin{equation}\label{L-amp-2}
 \langle V_1 V_2 \dots V_L \rangle = {\cal Z}_{\rm qu} \cdot {\cal Z}_{\rm cl} =
{\cal Z}_{\rm qu} \cdot \sum_{\{X_{\rm cl}\}} \exp(-S_{\rm cl}),
\end{equation}
where $X_{\rm cl}$ is the solution to the classical equation of motion.
The classical part is formally characterized as decomposable
part and physically gives instanton of worldsheet nature, via the
exchange of intermediate string. That gives
intuitive understanding via the `area rule', where 
the area corresponds to one, which intermediate string sweeps.

In the three-point amplitude, the summation of the classical action 
$\sum_{\{X_{\rm cl}\}} \exp(-S_{\rm cl})$  becomes the theta 
function~\cite{CIMYukawa}, where $S_{\rm cl}$ corresponds to 
the triangle area.
When we exchange $\tau$ and $J$ as (\ref{T-dual}) in the magnetized models, 
the Yukawa coupling (\ref{3pt}) corresponds to the following expansion 
\begin{equation} \label{intersec3pt} \begin{split}
  y_{i j \bar k} & = \jtheta{{M_2 k - M_3 j + M_2 M_3 l \over M_1 M_2
      M_3} \\ 0} \left(0,iM_1 M_2 M_3 A/(4 \pi^2 \alpha') \right)
  \\
    & = \sum_{n \in \Z} \exp\left[-\frac{M_1 M_2 M_3 A}{4\pi \alpha'}
  \Big({M_2 k - M_3 j + M_2 M_3 l \over M_1 M_2 M_3}+n \Big)^2\right], 
\end{split} \end{equation}
by using the definition (\ref{jacobitheta}).
We have neglected the antisymmetric tensor component $B$. 
The exponent corresponds the area (divided by $4 \pi \alpha'$) of 
possible formation of triangles and the one with $n=0$
corresponds to the minimal triangle.
Recall that the theta function part depends only 
$\tau$ and $J$ in magnetized and intersecting D-brane models, 
respectively.

We have omitted the normalization factor, corresponding to the quantum
part ${\cal Z}_{\rm qu}$. It is obtained by comparing the coupling
(\ref{intersec3pt}) with (\ref{Yukawa}). We find the factor
\begin{equation}
 2^{-9/4} \pi^{-3} e^{\phi_4/2} \prod_{d=1}^3 \left(\Im \tau_d {M_1^{(d)}
    M_2^{(d)} \over M_3^{(d)}}  \right)^{1/4}, 
\end{equation}
in the magnetized brane side corresponds to 
\begin{equation}
 {\cal Z}_{\rm qu} = (2 \pi)^{-9/4} e^{\phi_4/2} \prod_{d=1}^3 \left((\Im
 J_d)^2 {\theta_1^{(d)} \theta_2^{(d)} \over \theta_3^{(d)}}
 \right)^{1/4}, 
\end{equation}
in the intersecting brane side. We obtain the four dimensional dilaton
$\phi_4 = \phi_{10} - \ln |\Im \tau_1
\Im \tau_2 \Im \tau_3|$ from the ten dimensional one $\phi_{10}$, 
which is related with $g_{\rm YM}$ as  $g_{\rm YM} =
e^{\phi_{10}/2}\alpha^{\prime 3/2}$.
The vacuum expectation value of the dilaton gives gauge coupling $
e^{\langle \phi_4 \rangle /2} = g$.
In this case, the factor containing the angles is a leading order approximation of the ratio of Gamma function
\begin{equation}
 { \Gamma(1-\theta_1) \Gamma(1-\theta_2) \Gamma(\theta_3) \over \Gamma(\theta_1) \Gamma(\theta_2) \Gamma(1-\theta_3)} \simeq {\theta_1 \theta_2 \over \theta_3},
\end{equation}
valid for small angles. 
Therefore, the three-point couplings coincide each other between 
magnetized and intersecting D-brane models.
That is the observation of \cite{CIM}.

\begin{figure}[t] \begin{center}
\includegraphics[height=2cm]{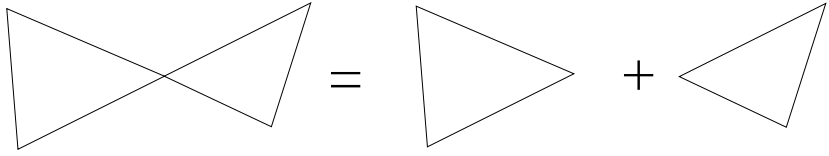}
\caption{Area of polygon, responsible for the classical part exponent,
  is decomposed in terms of those of three point functions. }
\label{f:areap}
\end{center} \end{figure}

Now, let us consider the four-point coupling 
of intersecting D-brane model corresponding to 
the left figure of Fig.~\ref{f:areap}.
The four-point amplitude is written as (\ref{L-amp-2}), 
where the classical action corresponds to 
the area of the left figure.
However, that can be decomposed into two triangles 
like the right figure, that is, 
the classical part can be decomposed into two parts, 
each of which corresponds to the classical part of 
three-point amplitude, i.e. 
\begin{equation}\label{decomp4-3}
\exp (-S_{\rm cl}^{(4)})=
\exp (-{S}_{\rm cl}^{(3)}) \exp (-{S'}_{\rm cl}^{(3)}),
\end{equation}
where $S_{\rm cl}^{(4)}$ corresponds to the area 
of the left figure of Fig.~\ref{f:areap} and 
${S}_{\rm cl}^{(3)}$ and ${S'}_{\rm cl}^{(3)}$ 
correspond to the triangle areas of the 
right figure.

On the other hand, our results in the previous sections show that 
the four-point
coupling in the magnetized model is also expanded as (\ref{sdecomp-2}).
Each of theta functions in  (\ref{sdecomp-2}) corresponds to 
the classical parts of the three-point couplings 
in the intersecting D-brane models.
This relation corresponds to the 
above decomposition (\ref{decomp4-3}).
Thus, the theta function parts of the four-point couplings, 
i.e. the classical part, coincide each other between 
magnetized and intersecting D-brane models.  
That means that the holomorphic complex structure, $\tau$, 
dependence of the four-point couplings 
in the magnetized brane models is the same as 
the holomorphic K\"ahler moduli $J$ dependence in the intersecting 
D-brane models, since the theta function part in 
the magnetized (intersecting) D-brane models depends 
only on $\tau$ ($J$).
The other part in the magnetized brane models corresponds 
to normalization factors $N_M$.
When we take a proper normalization, these 
factors also coincide.

\section{Conclusions}
We have calculated three-point and higher order couplings 
of four-dimensional effective field theory 
arising from dimensional reduction of magnetized brane models.
We have found that higher order couplings are written 
as products of three-point couplings.
This behavior is the same as higher order 
amplitudes of CFT, that is, higher order amplitudes 
are decomposed as products of three-point 
amplitudes in intersecting D-brane models.
Our results on higher order couplings 
would be useful in phenomenological applications.
Numerical analysis on higher order couplings is also possible.

\subsection*{Acknowledgement}
H.~A. is supported by the Grant-in-Aid for the Global COE Program 
``Weaving Science Web beyond Particle-matter Hierarchy'' from the 
Ministry of Education, Culture, Sports, Science and Technology of Japan. 
K.-S.~C. and T.~K. are supported in part by the Grant-in-Aid for 
Scientific Research No.~20$\cdot$08326 and No.~20540266 from the 
Ministry of Education, Culture, Sports, Science and Technology of Japan.
T.~K. is also supported in part by the Grant-in-Aid for the Global COE 
Program "The Next Generation of Physics, Spun from Universality and 
Emergence" from the Ministry of Education, Culture,Sports, Science and 
Technology of Japan.

\end{document}